\documentstyle[cjaa209,twoside,epsf]{article}

\begin{document}

\title{ Multifrequency Radiation\\
of Extragalactic Large-Scale Jets }

\author{\L ukasz Stawarz  }

\inst{
Obserwatorium Astronomiczne, Uniwersytet Jagiello\'{n}ski, ul. Orla 171, 30-244 Krak\'{o}w, Poland\\
}

\email{stawarz@oa.uj.edu.pl}

\markboth{\L. Stawarz}
{Multifrequency Radiation of Extragalactic Large-Scale Jets }

\pagestyle{myheadings}

\date{Received~~2003~~~~~~~~~~~~~~~ ; accepted~~2004~~~~~~~~~~~~~~ }

\baselineskip=18pt

\begin{abstract}
\baselineskip=18pt
Large-scale extragalactic jets, observed to extend from a few to a few hundred kiloparsecs from active galactic nuclei, are now studied over many decades in frequency of electromagnetic spectrum, from radio until (possibly) TeV $\gamma$ rays. For hundreds of known radio jets, only about 30 are observed at optical frequencies. Most of them are relatively short and faint, with only a few exceptions, like 3C 273 or M 87, allowing for detailed spectroscopic and morphological studies. Somewhat surprisingly, the large-scale jets can be very prominent in X-rays. Up to now, about 30 jets were detected within the $1 - 10$\,keV energy range, although the nature of this emission is still under debate. In general, both optical and X-ray jet observations present serious problems for standard radiation models for the considered objects. Recent TeV observations of M 87 suggest the possibility of generating large photon fluxes at these high energies by its extended jet.\\
In this paper we summarize information about multiwavelength emission of the large-scale jets, and we point out several modifications of the standard jet radiation models (connected with relativistic bulk velocities, jet radial stratification and particle energization all the way along the jet), which can possibly explain some of the mentioned puzzling observations. We also comment on $\gamma$-ray emission of the discussed objects.
\end{abstract}

\section{Introduction}
\label{sect:Int}
The very first extragalactic large-scale jet was discovered in optical. This was a `curious straight ray' emanating from the nucleus of an active galaxy M 87 (Curtis~\cite{cur18}), which was called a `jet' for the first time by Baade and Minkowski (\cite{baa54}) and detected at radio frequencies much later. Also the second known extragalactic jet, connected with the quasi-stellar object 3C 273, was found in optical at first (Schmidt~\cite{sch63}). One can note, that the large-scale jet in the closest radio galaxy Centaurus A was observed at X-rays before its radio structure was revealed (Schreier et al.~\cite{sch79}). However, it was the development of {\it Very Large Array} ({\it VLA}) and {\it Very Long Baseline Interferometry} ({\it VLBI}) radio techniques, which enabled the first systematic observations of many jets associated with distant radio galaxies and other kinds of Active Galactic Nuclei (AGNs). Since then, studying extragalactic jets was related predominantly to radio observations, which could follow them from tens-of-kiloparsec down to parsec scales. Until now, almost one thousand radio jets (accordingly to the morphological definition by Bridle and Perley~\cite{bri84}) in more than six hundred radio sources were discovered (see Liu and Zhang~\cite{liu02}). For a long time, only a few of these objects were known to possess optical or X-ray counterparts. Recently, the situation started to change after the optical {\it Hubble Space Telescope} ({\it HST}) and {\it Chandra X-ray Observatory} ({\it CXO}) were launched\footnote{See the web pages \texttt{http://home.fnal.gov/\~{}jester/optjets} by S. Jester\\ and \texttt{http://hea-www.harvard.edu/XJETS} by D. Harris.}.

Extragalactic large-scale jets are the largest physically connected structures in the Universe, with linear sizes reaching even megaparsecs. Although widely studied in radio frequencies over the last three decades and considered to play an essential role in unification theories for AGNs, they are still superficially known objects. Among the other issues, the exact jet composition, velocity and magnetic internal structure, or finally processes controlling energy dissipation and jet interactions with the surrounding (galactic and intergalactic) medium, are still under debate. For a wide discussion regarding these problems one can consult a review by Begelman et al.~\cite{beg84}, one of the more recent monographs (e.g., Hughes~\cite{hug91}), or finally some conference proceedings (e.g., Ostrowski et al.~\cite{ost97}). As a short introduction let us only note that extragalactic large-scale jets are usually assumed to be composed of fully ionized, collisionless and electrically neutral plasma, which is underdense with respect to intergalactic medium, containing ultrarelativistic nonthermal electrons and magnetic field frozen-in to the plasma.

Recent progress in studying extragalactic large-scale jets at high radiation frequencies substantially lighted up discussions regarding the physical conditions within these objects. Detailed optical and X-ray observations performed by {\it HST} and {\it CXO} with high spatial resolution and high sensitivity already gave us a number of new, in many aspects unexpected and puzzling results. Here we will try to briefly summarize these new observations, and to discuss what they can tell us about radiating particles (their spatial localization and spectral energy distribution), and, more generally, about the physical conditions within the large-scale jets (is there any \emph{in situ} particle re-acceleration within the jet flow? if yes, what is its nature?). In this context, we will briefly mention theoretical models proposed in literature till now, which try to explain optical and X-ray data. Finally, we will also discuss an exciting possibility of observing extragalactic large-scale jets at very high $\gamma$-ray frequencies, open for us thanks to modern {\it Imaging Atmospheric Cherenkov Telescopes} ({\it IACTs}) or planned space missions like {\it $\gamma$-ray Large Area Space Telescope} ({\it GLAST}). Obviously, any discussion should start with a review of previous and recent radio observations.

\section{Radio emission}
\label{sect:Rad}
Extragalactic jets as observed in radio frequencies are more or less narrow beams emanating from compact central cores of radio-loud AGNs -- radio galaxies, quasars, Seyfert galaxies or BL Lac objects -- and exhibiting a complexity and variety of morphology and physical properties. Each radio-loud AGN is supposed to produce a pair of jets flowing with at least mildly relativistic bulk velocities in opposite directions, being called `a (main) jet' (the one approaching the observer) and `a counterjet' (the receding one). According to our current knowledge, jets continuously transport energy in a form of bulk kinetic (or electromagnetic) energy from active centers to the kpc up to Mpc scales. There the carried nonthermal plasma loses collimation, and being eventually turned back by the ambient medium, forms two extended radio `lobes' at both sides of the parent galaxy. In a few recent cases of powerful and extremely extended radio sources, two pairs of aligned and hosted by the same galaxy double-lobed structures of different sizes were discovered (e.g., Schoenmakers et al.~\cite{sch00}, Saripalli et al.~\cite{sar02,sar03}), showing that the jet activity can be intermittent (see in this context Clarke et al.~\cite{cla92}, Kaiser et al.~\cite{kai00}). During the jet's active period, a small part of jet kinetic energy is converted to energization of nonthermal particles producing synchrotron (and also inevitably inverse-Compton) emission along the flow. The most powerful jets finish with bright `hot spots', where strong shocks form and convert the jet kinetic energy into random internal energy of the shocked plasma. This paper does not deal with multiwavelength emission of the radio lobes or of the hot spots, although it should be mentioned that by studying them one provides important constraints on jet parameters. For example, an analysis of the extended radio regions surrounding powerful sources suggests that the power of their jets is at least $10^{47}$\,erg s$^{-1}$ (Rawlings and Saunders~\cite{raw91}). In this context, new X-ray observations of lobes and hot spots (see, e.g., Brunetti~\cite{bru02b}) are of particular interest.

Identification of extragalactic large-scale radio jets can be difficult because of the aforementioned variety of jet morphology and observational limitations. Sometimes one does not see a continuous beam, but rather a train of `knots' (brightenings) along the expected jet path, probably due to the much lower surface brightness of the inter-knot structures. Knotty morphology of the extragalactic jets, coexisting with their good collimation over many decades in length scale, is one of the most striking and important characteristics of these objects. However, jets can also be strongly bent and even suggestively disrupted when leaving the host galaxy, particularly if localized in rich cluster medium. Finally, the jets can be invisible although their by-products, like hot spots or extended lobes, are clearly detected. This is the case for most of the counterjets, especially in powerful sources. In general, lobes and hot spots at both sides of the parent active galaxy (if present) are much more symmetric in size and radio power than the jet and the counterjet.

Most of the radio jets are one-sided close to their parent objects, i.e. at the pc scales. Further out from the nucleus, at kpc scales, jets in weak sources become gradually or suddenly two-sided, while most of the jets in powerful sources propagate through all of their tens- to hundreds-of-kpc lengths with high jet-counterjet brightness asymmetry. This behavior seems to be connected with an overall morphology of the radio source, which led Fanaroff and Riley (\cite{fan74}) to divide radio galaxies into two classes. Radio galaxies of type I (FR Is) are therefore characterized as weak sources, with total power at $178$\,MHz being $P_{178 \, {\rm MHz}} < 2 \cdot 10^{25}$\,W Hz$^{-1}$, which are edge-darkened (i.e. for which the radio intensity decreases away from the core), lacking hot-spots and possessing two-sided large-scale radio jets (with eventually small jet-counterjet brightness asymmetry). Radio galaxies of type II (FR IIs, `classical doubles') are powerful, $P_{178 \, {\rm MHz}} > 2 \cdot 10^{25}$\,W Hz$^{-1}$, edge-brightened, with an entirely one-sided radio jet and with two bright hot spots at the outflow terminal points. There are some other differences between these two classes, for example jets in FR IIs are generally better collimated (almost cylindrical), less bended and more knotty (i.e. exhibiting higher knot-to-interknot contrast) than the jets in FR Is. Also, opposite to the FR I radio galaxies, extended radio lobes in FR IIs dominate the total radio power. For this reason, the jets in the later sources, although more powerful than the weak FR I jets, are more difficult to detect. Well studied examples of these two types of radio sources are FR I radio galaxy 3C 31 (e.g., Laing and Bridle~\cite{lai02}) and FR II object Cygnus A (e.g., Carilli and Barthel~\cite{car96}).

Jets in radio-loud quasars are similar to the jets in FR II radio galaxies, but are more prominent in total radio luminosity of a system and, therefore, are being detected more often. On the other hand, large-scale radio structures in BL Lac objects are supposed to resemble those in FR Is. The correspondence between quasars and FR II radio galaxies as well as between BL Lacs and FR Is, is one of the basic ideas of the unification scheme for radio-loud AGNs (e.g., Urry and Padovani~\cite{urr95}). However, it is still required to be verified carefully by means of multifrequency and multi-scale observations. It should be noted that there are some radio sources with a BL Lac nucleus and FR II large-scale radio morphology (e.g., 3C 371). Recently Blundell and Rawling (\cite{blu01}) discovered also the $300$\,kpc-long FR I jet in the quasar E1921+643 classified previously as radio-quiet. This suggests that at least some of the radio-quiet quasars may have FR I large-scale jet morphology. The radio-loud/radio-quiet dichotomy, and, in general, the accuracy of the jet unification scheme (exploiting predominantly kinematical factors like jet inclination and velocity), is one of the most widely discussed topics recently. The same is true for understanding the FR I - FR II boundary, especially after the discovery of a few HYbrid MOrphology Radio Sources (HYMORS), which exhibit an FR I radio structure on one side of the nucleus, and an FR II on the other (Gopal-Krishna and Wiita~\cite{gop00}). The presence of such objects supports an `extrinsic' explanation for the FR physical separation, ascribing it to the jet interaction with the surrounding medium controlled by its power, and opposes `intrinsic' explanations, which consider differences in the jet composition, spin of the central black hole or accretion process.

Jet-counterjet brightness asymmetry, being correlated with radio luminosity of the source and usually changing along the flow, can be used to estimate jet bulk velocity, or even to study the jet velocity structure. A natural explanation for such asymmetry is a relativistic beaming, which can amplify or de-amplify emission produced by a relativisticaly moving source accordingly to $L = (\delta^3/\Gamma) \, L'$, where $L$ is the observed synchrotron luminosity of the continuous, steady jet, $L'$ is the appropriate intrinsic luminosity, $\Gamma$ is a jet bulk Lorentz factor and $\delta = 1/ \Gamma (1 - \beta \cos \theta)$ is the Doppler factor for the jet moving with the velocity $\beta = \sqrt{1 - \Gamma^{-2}}$ with an inclination $\theta$ to the line of sight (e.g., Lind and Blandford~\cite{lin85}, Sikora et al.~\cite{sik97}). Radio observations suggest, that at parsec distances from the active centers all jets in radio galaxies and radio loud quasars are highly relativistic, with $\Gamma \sim 3 - 10$, independently on their radio luminosity and large-scale structure classification (Giovannini et al.~\cite{gio01}). At larger scales, two-sideness of FR I jets suggest only mildly- or even sub-relativistic velocities (e.g., Laing et al.~\cite{lai99}). On the other hand, detailed {\it VLA} studies of the FR II sources imply that jet speeds in these objects remain relativistic up to the flow termination in hot-spots, however with bulk Lorentz factors not greater than $\Gamma \sim 3$ (Wardle and Aaron~\cite{war97}). Thus, jet deceleration between pc- and kpc-scales seems to take place, being in addition stronger in weak FR I sources than in FR IIs.

A great deal of theoretical effort was devoted to explain how FR I jets can decelerate at first kiloparsecs from the jet base, without loosing collimation and avoiding efficient energy dissipation to radiating particles. This can probably be understood in terms of significant entraining of the ambient (galactic) medium by such jets, resulting in changing the character of their flow from an initially relativistic and supersonic to the sub-relativistic, highly turbulent and subsonic (Begelman~\cite{beg82}, de Young~\cite{dey86}, ~\cite{dey93}, Bicknell~\cite{bic94}, Komissarov~\cite{kom94}). The situation in FR IIs is still an open question, and it is possible that these jets propagate with much higher $\Gamma$ than the radio observations suggest (see next Sects.). This confusion arises from the fact that the only direct information we have about these objects is obtained from their non-thermal radiation, which usually cannot give us definite constraints on the jet parameters, especially if observations are limited to the narrow (radio) frequency range. As a result, it is not clear if the jet-counterjet brightness asymmetry measurements can give us reliable estimates of the large-scale jet bulk velocities, as it can be seriously affected by, for example, the interaction of the jet with the surrounding medium. Such interactions can create velocity structure across the jet flow, and also stimulate particle acceleration processes. These effects can significantly influence jet-counterjet radio brightness asymmetry, and in general the whole jet radiative output, but are difficult to quantify (Komissarov~\cite{kom90}, Stawarz and Ostrowski~\cite{sta02b}).

Radio emission of extragalactic large-scale jets is a polarized power-law continuum, $L_{\nu} \propto \nu^{- \alpha}$, with high linear polarization reaching $40 \%$, or sometimes even more than $50 \%$, and with a radio spectral index being typically $\alpha_{\rm R} \sim 0.5 - 0.8$. The observed radio luminosities of jets change from $10^{38}$\,erg s$^{-1}$ for weak sources, up to $10^{44}$\,erg s$^{-1}$ for the most luminous ones. A natural explanation of this emission is a synchrotron radiation of ultrarelativistic electrons with a nonthermal energy distribution $n'_{\gamma}$, spinning in a tangled magnetic field of intensity $B$. From the standard synchrotron theory, in which every electron with the Lorentz factor $\gamma$ radiates mostly at the critical frequency $\nu' = (e / 4 \pi \, m_{{\rm e}} c) \, B \, \gamma^2$, one has (in the jet commoving frame denoted by primes) $L'_{\nu'} d\nu' = m_{\rm e} c^2 \, | \dot{\gamma} | \, V' \, n'_{\gamma} d\gamma$, where a rate of synchrotron cooling is $m_{\rm e} c^2 \, | \dot{\gamma} | = {4 \over 3} \, c \sigma_{\rm T} \, U'_{\rm B} \gamma^2$, magnetic field energy density is $U'_{\rm B} = B^2 / 8 \pi$ and $V'$ is the volume of the emitting region. Therefore, the electron energy distribution required to produce the observed radio emission has to be a power-law, $n'_{\gamma} \propto \gamma^{-s}$, with the spectral index $s = 2 \alpha_{\rm R} + 1 \sim 2 - 2.6$, and with low (`l') and high (`h') energy cut-offs related to the maximum and minimum observed synchrotron frequency by $\gamma_{{\rm l/h}} \sim 0.1 \, B_{-4}^{-1/2} \, (\nu'_{{\rm min/max}})^{1/2}$, where $B_{-4} \equiv B / 10^{-4}$\,G and $\nu'$ is given in Hz (note that the radio observations usually do not cover the whole range of the electron synchrotron emission). Such electron spectra are obtained in a natural way in Fermi-type processes (first-order shock acceleration or second-order stochastic acceleration), in which relativistic particles undergoing multiple scatterings within turbulent magnetized plasma increase their energies and can form a power-law energy distribution.

One cannot derive the magnetic field intensity solely from the radio synchrotron emission. The problem is usually overcome by applying assumption of energy equipartition between relativistic particles and the jet magnetic field, which minimizes the total pressure (energy) of the synchrotron source with given volume and luminosity (e.g., Miley~\cite{mil80}). However, in order to use this procedure, one should know the composition of the jet relativistic matter, in particular the ratio of the relativistic proton pressure to the radiating electron pressure, and to evaluate the minimum electron energy, crucial in deriving the electron energy density for $s > 2$ ($\alpha_{\rm R} > 0.5$). Usually, when computing the magnetic field equipartition value, the pressure of relativistic protons is neglected, and the electron minimum energy is related to the lowest observed radio frequency, $\nu_{{\rm min}} =\delta \, \nu_{{\rm min}}' \sim 10$\,MHz, being then $\gamma_{{\rm l}} \sim 3 \cdot 10^2 \, B_{-4}^{-1/2} \, \delta^{-1/2}$. Both these additional assumptions are not necessarily correct, as indicated by the new X-ray observations of the quasar jets: the former one seems to be violated in the case of small scale blazar sources (Sikora and Madejski~\cite{sik00}), while the latter one - probably, but not necessarily -- in the case of their large-scale counterparts observed by {\it CXO} (see next Sects.). Whatever the case, the equipartition values of the magnetic field in the large-scale extragalactic jets, being typically $B_{{\rm eq}} \sim 10^{-4}$\,G, should be considered as rough order-of-magnitude estimates of real values.

Also the spatial structure of the jet magnetic field remains largely unknown. Possibly it is tangled on a small scale, and eventually partially ordered on a large scale due to its own regular topology, but also due to shearing and/or compression of the jet medium (Laing~\cite{lai80}). In general, in the absence of the Faraday effects, polarization maps imply that within powerful large-scale jets the magnetic field is on average parallel to the jet axis. In the case of FR Is the situation is more complex, with the magnetic field changing its configuration from parallel to transverse along the flow, across the flow (in the sense that it is aligned to the jet axis near the jet edges and transverse at the jet center), or even in some cases exhibiting pure transverse behavior. Although it is not easy to guess the realistic three-dimensional configuration of the magnetic field from the polarization maps alone, such observations are usually interpreted as evidence for the jet internal structure, consisting of a fast central spine characterized by the transverse magnetic field (due to shock compression?), and surrounded by the boundary layer with the magnetic field aligned parallel to the jet axis due to shearing effects arising from interaction with the surrounding medium. A different interpretation involving helical magnetic field -- in the context of the pc-scale jets -- was proposed by, e.g., Gabuzda (\cite{gab99}). Spine-boundary shear layer polarization structures appended with kinematic interpretation of the jet-counterjet brightness asymmetry are consistent with the model of decelerating FR I jet, for which an initially relativistic flow is visible at the beginning only due to its slower boundary layer, but after the jet decelerates, its emission starts to be dominated by the spine component (Laing~\cite{lai96}). One should note, that particle acceleration processes acting in both jet components could differ substantially.

There are several models of particle acceleration processes considered in the literature in the context of extragalactic large-scale jets, for example the ones involving shock waves within the flow. This class of models identifies knots with places of strong shocks, where ultrarelativistic particles are accelerated via the first-order Fermi process and form power-law energy distributions with spectral indices $s \sim 2$ in a non-relativistic case (e.g., Blandford and Eichler~\cite{bla87}; for an ultrarelativistic case see, e.g., Ostrowski and Bednarz~\cite{ost02}). It was proposed by Rees (\cite{ree78}), that such strong shocks could form due to intrinsic velocity irregularities within the jet flow itself. Alternatively, shocks can form in the knot regions due to interactions of the jet plasma with dense clouds of the external matter (Blandford and Konigl~\cite{bla79}), due to jet reconfinement by the ambient medium (e.g., Sanders~\cite{san83}), or finally due to large-scale instabilities within the jet flow\footnote{If strong enough, such instabilities can also alter the large-scale jet morphology. In this context Eilek et al. (\cite{eil02}) proposed an alternative to the FR classification of extragalactic radio sources, which exploits ability of jet flows to development and eventual saturation of the disruptive large-scale instabilities.} (e.g., Bicknell and Begelman~\cite{bic96}). An alternative and/or supplementary class of models involve the second-order Fermi acceleration of relativistic particles on turbulent MHD waves, created within the jet flow by inevitable Kelvin-Helmholtz instabilities (Pacholczyk and Scott~\cite{pac76}, Lacombe~\cite{lac77}, Eilek~\cite{eil79}, Bicknell and Melrose~\cite{bic82}). Under several assumptions, connected, for example, with poorly understood processes determining spectrum and structure of the turbulence, these models can explain some aspects of the observed jet radio emission. One of the consequences of these models is the fact that as the Kelvin-Helmholtz instabilities are excited predominantly at the jet boundaries, where they form turbulent boundary layer by cascading to the shorter wavelength (e.g., Birkinshaw~\cite{birk91}), dissipation of the jet kinetic energy to radiating particles should take place predominantly at the jet edges (e.g., Eilek~\cite{eil82}; cf. Bicknell and Melrose~\cite{bic82}). As a result, one should expect limb-brightening of the jets, what is in fact the case for some radio sources (if only observed with a sufficient spatial resolution), like FR I jet in M 87 or FR II jet in 3C 353 (Owen et al.~\cite{owe89} and Swain et al.~\cite{swa98}, respectively). However, additional physical processes can complicate such a simple picture, for example a turbulent mixing of the spine and the boundary layer medium. Finally, another, and at the moment hardly explored, possibility for particle acceleration within the large-scale jets is connected with magnetic reconnection processes (e.g., Birk and Lesch~\cite{bir00}).

\section{Optical emission}
\label{sect:Opt}
Before the {\it HST} mission, only a few extragalactic large-scale jets were observed at optical photon energy range. There were the jets in the FR I radio galaxies M 87 (Curtis~\cite{cur18}) and 3C 66B (Butcher et al.~\cite{but80}), in the quasar 3C 273 (Schmidt~\cite{sch63}), and in the BL Lac object with a hybrid radio morphology PKS 0521-365 (Danziger et al.~\cite{dan79}). Isolated optical knots related to the jet radio structure were also observed in the FR II source 3C 277.3 (Miley et al.~\cite{mil81}), and in the intermediate FR I/FR II radio galaxy NGC 6251 (Keel~\cite{kee88}). In addition, detection of a weak optical jet in the FR I object 3C 31 was reported by Butcher et al. (\cite{but80}), but confirmed only recently (Croston et al.~\cite{cro03}). Finally, bright optical emission considered previously as a manifestation of the hot spot in the FR II source 3C 303 (e.g., Keel~\cite{kee88}), was later recognized as being caused by the knot activity (Meisenheimer et al.~\cite{mei97}). In general, in the case of FR II sources, discrimination between the hot-spot and knot origin of the optical emission can be confusing sometimes. This, however, is important, as the physical conditions in hot-spots can be much different from the ones within the jets themselves. Another observational difficulty that should be mentioned is the requirement of very careful extraction of the host galaxy starlight in order to reveal the presence of short optical jets emanating from the galactic nuclei. Also, isolated jet brightenings -- knots -- can be misled with background/foreground optical objects, like galaxies or intergalactic gas clouds. Therefore, in addition to the apparent coincidence between optical features with the radio jet structure, additional spectral information is sometimes required in order to confirm the presence of the optical counterpart to the radio jet. Such observational limitations have been partly overcome by the {\it HST}, starting from its first, unexpected detection of a short (projected length $\sim 0.3$\,kpc) optical jet in the FR I radio galaxy 3C 264 (Crane et al.~\cite{cra93}).

Until now, about 30 optical jets in extragalactic sources are known, plus several non-discussed here candidate sources. Clear, elongated and continuous optical jet-like structures, or only isolated optical knots corresponding to the radio ones, are observed in different types of radio sources. There are optical jets in the quasars: 3C 273 (e.g., R\"oser and Meisenheimer~\cite{ros91}, Bahcall et al.~\cite{bah95}, Jester et al.~\cite{jes01}), 3C 212 and 3C 245 (Ridgway and Stockton~\cite{rid97}), 3C 380 (de Vries et al.~\cite{dev97}), PKS 0637-752 (Schwartz et al.~\cite{sch00}), 3C 279 (Cheung~\cite{che02}), PKS 1136-135, PKS 1150+497 and PKS 1354+195 (Sambruna et al.~\cite{sam02}), 3C 179, 3C 207, 3C 345, PKS 1642+690, PKS 1928+738, PKS 2251+134 (Sambruna et al.~\cite{sam04b}); and in the FR II radio galaxies: 3C 371 (Nilsson et al.~\cite{nil97}, Scarpa et al.~\cite{sca99}), 3C 277.3 (Miley et al.~\cite{mil81}), 3C 303 (Meisenheimer et al.~\cite{mei97}) and 3C 346 (Dey and van Breugel~\cite{dey94}). There are also the optical jets in the intermediate FR I/FR II objects NGC 6251 and 3C 15 (Keel~\cite{kee88} and Martel et al.~\cite{mar98}, respectively), as well as in the aforementioned hybrid morphology radio source PKS 0521-365 (Danziger et al.~\cite{dan79}, Scarpa et al.~\cite{sca99}). Optical jets are also being observed in a number of FR I radio galaxies, like M 87 (e.g., Meisenheimer et al.~\cite{mei96}, Perlman et al.~\cite{per99},~\cite{per01}), 3C 66B (Butcher et al.~\cite{but80}, Macchetto~\cite{mac91}), 3C 31 (Butcher et al.~\cite{but80}, Croston et al.~\cite{cro03}), 3C 264 (Crane et al.~\cite{cra93}, Lara et al.~\cite{lar99}), 3C 78 (Sparks et al.~\cite{spa95}), PKS 2201+044 (Scarpa et al.~\cite{sca99}), B2 0755+37 and B2 1553+24 (Parma et al.~\cite{par03}) and in the Seyfert galaxy 3C 120 (Hjorth et al.~\cite{hjo95}). The projected lengths of these jets, located always on a main-jet side, range from a few hundred parsecs up to a few kiloparsecs in cases of FR I sources, and from a few up to a few tens of kiloparsecs in the cases of quasars and FR IIs. However, the optical jets are always shorter than their radio counterparts. They are also narrower, better collimated (almost cylindrical, with a beam radius $\sim 0.1 - 1$\,kpc), and more knotty (i.e. with higher knot-to-interknot brightness contrast) than the related radio structures. No counterjet to any of the aforementioned objects was detected in optical (but see Fraix-Burnet~\cite{fra97}). Interestingly, in 3C 66B and 3C 264 a filamentary, edge-brightened (`double-stranded') optical jet structure was found (Macchetto et al.~\cite{mac91}, Crane et al.~\cite{cra93}, respectively)

Optical emission observed from the discussed objects is a featureless power-law continuum. The optical spectral index ranges from $\alpha_{\rm O} \sim 0.6$ (e.g., 3C 371) up to $\geq 1.6$ (in 3C 66B, PKS 0521-365, 3C 279, 3C 346), and is in general larger than the radio one. Radio-to-optical power-law slope ranges from $\alpha_{{\rm R-O}} \sim 0.6$ (e.g., 3C 264) up to $\sim 1$ (for 3C 212, 3C 245). In some cases, linear polarization of the optical emission is measured on the level of ten percent and more (M 87, 3C 273, 3C 120 3C 277.3). This fact, together with an almost one-to-one correspondence in the knotty morphology between the radio jets and their optical counterparts, strongly argues for the synchrotron origin of the jet optical photons, invalidating alternative possibilities like Thomson scattering of AGN anisotropic radiation, thermal bremsstrahlung of surrounding hot clumped gas, or emission of jet-induced star formation regions. Therefore, for the considered jets the radio power-law continuum extends with a spectral index $\alpha_{\rm 1} < 1$ up to some break frequency $\nu_{{\rm br}}$ placed around optical energy range, where it breaks to $\alpha_{\rm 2} > 1$. As a result, most of the radiative power is emitted at frequencies $\nu \sim \nu_{{\rm br}}$. One should note, however, that in most cases the exact location of the break frequency is not well constrained and, in addition, there are indications that it varies significantly from object to object: among the jets observed in optical and also X-rays $\nu_{{\rm br}}$ can be placed somewhere between infrared ($\nu_{{\rm br}} \geq 10^{12}$\,Hz) up to ultraviolet ($\nu_{{\rm br}} \leq 10^{16}$\,Hz) frequencies. This manifests itself by the large dispersion of jet optical spectral indices.

The most remarkable characteristic of the optical extragalactic jets is that both radio-to-optical spectral indices, as well as equipartition values of the jet magnetic field are constant, or only very smoothly changing along the jet. This behavior was noted in almost every case (PKS 0521-365, 3C 264, 3C 66B), but most clearly in 3C 273 and M 87. In the latter source, both $\alpha_{\rm O}$ and $\alpha_{{\rm R-O}}$ oscillate around the mean values $\sim 0.9$ and $\sim 0.7$, respectively. Smooth changes of these spectral indices are not correlated with each other, although in some jet regions subtle anti-correlation was noted, as well as correlation between $\alpha_{\rm O}$ and the jet brightness (Meisenheimer et al.~\cite{mei96}, Perlman et al.~\cite{per01}). For such physical conditions, the required Lorentz factor of electrons responsible for optical synchrotron emission, $\gamma \sim 10^6$, should be stable over $2$\,kpcs of the M 87 jet length. In addition, the considered object exhibits strong (up to $50 \%$) polarization all along the flow, also in the interknot regions, both in radio and optical, indicating a well-ordered magnetic field. Interestingly, the magnetic field configuration is much different when observed at these two photon energy domains, in the sense that optical polarization is stronger at the jet center, while the radio one is larger at the jet edges. Also, the magnetic field inferred from the optical observations reveals longitudinal configuration in the interknot regions but transverse within the bright knots, while it is always parallel to the jet axis when observed in radio (Perlman et al.~\cite{per99}). This suggests spine - boundary shear layer morphology, with higher energy emission concentrated within a (faster) central jet component.

The optical jet in 3C 273 looks different from the one in M 87, although some general similarities remain. Firstly, it is much longer, with a projected length $\sim 37$\,kpc. Similarly to M 87, the optical jet in 3C 273 is narrower and much better collimated than its radio counterpart. It is characterized by a series of bright knots confined to a cylindrical tube and linked by diffusive and continuous inter-knot emission. Such morphology was considered in terms of a fast moving central spine bright in optical, surrounded by a slow-moving radio cocoon (Bahcall et al.~\cite{bah95}). Both optical and radio polarisations reveal magnetic field being parallel (on average) to the jet axis. Similarly to the jet in M 87, optical and radio spectral indices, as well as radio-to-optical power-law slope, changes very smoothly along the jet. However, in this case $\alpha_{\rm O}$ does not fluctuate around the mean value, but instead decreases very slowly from $\sim 0.5$ in the inner jet regions up to $\sim 1.6$ in its outer parts, without a correlation to the optical surface brightness (Jester et al.~\cite{jes01}). In fact, there are large but local variations in the brightness distribution along the jet without any significant changes of the optical spectral index, but at the same time the optical flux remains globally constant along the jet while $\alpha_{\rm O}$ slowly declines. These observational facts have led Jester et al. (\cite{jes01}) to conclude that there is no evidence of efficient radiative cooling along a $30$\,kpc-long optical jet in 3C 273.

A lifetime of an ultrarelativistic electron with the Lorentz factor $\gamma$ -- for nonrelativistic jet velocities and $U_{\rm B} > U_{{\rm rad}}$ (where $U_{{\rm rad}}$ is energy density of radiation fields contributing to the electron inverse-Compton emission, and therefore to additional radiative losses) -- is $t_{{\rm life}} \sim \gamma / | \dot{\gamma} |_{{\rm syn}} \sim 10^3 \, \nu_{15}^{-1/2} \, B_{-4}^{-3/2} $\,yrs, where $\nu_{15} \equiv \nu / 10^{15}$\,Hz. With the equipartition magnetic field $B_{-4} = 1$ and $\nu_{15} = 1$ one obtains $t_{{\rm life, \, O}} \sim 10^3$\,yrs, what in other words limits propagation length of the optically emitting electrons $l_{{\rm O}} \leq c \, t_{{\rm life, \, O}} \sim 0.3$\,kpc. However, the projected length of the optical jets, as well as distances of optical knots from the active nucleus and from each other, are usually much longer than this scale, in some cases even more than by two orders of magnitude. Such discrepancy between optical jet length and the maximum propagation length of radiating electrons -- the so called `electron lifetime problem' -- was realized at the very beginning of optical jet observations. In addition, as mentioned above, radio-to-optical power-law slope is remarkably stable along the jets, and no signs of strong spectral ageing are observed. Although very knotty, some of the optical jets (like 3C 273 or M 87) clearly reveal the interknot optical emission, which is roughly constant over the spatial extent much longer than the above estimated $l_{{\rm O}}$ would allow. Therefore, even in a framework of the widely accepted scenario of shock acceleration taking place within bright knots, the optical jets are much too long, and the distances between the subsequent knots are too large to expect any optical synchrotron emission from the regions between them. As stressed by Meisenheimer et al. (\cite{mei96}), the shock-in-knot model alone cannot explain optical emission of the M 87 jet and some kind of permanent particle re-acceleration process acting along the jet body is required. It was further argued that the (equipartition) magnetic field controls smooth changes of the jet surface brightness, the optical spectral index, and even the acceleration efficiency.

Until today, different ideas explaining the electron lifetime problem were proposed. They can be divided into two main groups, namely either those involving some kind of particle re-acceleration within the jet flow, or the models avoiding \emph{in situ} energy dissipation to radiating particles. As extreme cases of the latter type, one can mention the models involving `loss-free' channels within the jets, forming due to the extremely low magnetic field intensity at the jet center (e.g., Owen et al.~\cite{owe89} proposed this idea to explain the limb-brightening radio structure of the M 87 jet) or due to a particular electromagnetic configuration of the jet flow (Kundt~\cite{kun89}, {\it `$E \times B$ drift'}). In such channels, ultrarelativistic electrons can propagate to large distances without significant synchrotron cooling and, then, eventually loose energy at some separate energy-loss sites, like the jet boundary layer with a significantly larger magnetic field. However, as discussed below, decreasing of the electron synchrotron cooling cannot solve the whole problem, as the inverse-Compton losses (due to, e.g., cosmic microwave background radiation or galactic radiation fields) can significantly limit the electron lifetime. A more widely considered explanation of the electron lifetime problem is related to the relativistic bulk velocities of the large-scale jets. In order to briefly discuss this possibility, let us review the following question connected with the optical jets in extragalactic radio sources: why are they so rare as compared to the radio jets? In particular, one would like to know what controls the `optical loudness' of the discussed objects. Is it some extrinsic reason, connected, e.g., with a character of the ambient medium controlling somehow jet radiative output (see, e.g., Fraix-Burnet~\cite{fra92}, also Nilsson et al.~\cite{nil97}), or intrinsic reason, like the age of the source for example?

Sparks et al. (\cite{spa95}) noted, that all optical jets known up to 1995 were hosted by galaxies with powerful and flat radio spectrum active nuclei. This fact, together with one-sideness of the discussed objects, their good collimation and shorter extensions as compared to the radio counterparts, led to the conclusion that relativistic beaming plays a substantial role in these jets. Sparks et al. concluded that the large-scale extragalactic jets with detected optical emission are significantly beamed. As the relativistic effects amplify the observed luminosity of the objects moving at an angle $\theta \leq 1 / \Gamma$ to the line of sign, a small number of detected optical jets is expected from the sample of randomly oriented ones with $\Gamma \gg 1$. Sparks et al. found also that the sources possessing optical jets have smaller projected linear sizes of the radio structures than the sources without optical jets but with comparable total radio power. As the projected linear size of the radio source depends crucially on the jet orientation, but not its total radio power, this result seems to be consistent with the beaming hypothesis. Let us note, however, that some new observations of radio sources with prominent optical jets do not fit well within the above picture. For instance, 3C 15 does not posses the bright unresolved nucleus, the optical jet in 3C 273 is not much shorter than the radio one, while B2 0775+37 and B2 1553+24 are very extended as compared to radio galaxies with similar total radio power. On the other hand, some other new observations support beaming hypothesis well. For example, Sparks et al. (\cite{spa00}) showed, that four of the nearest FR I galaxies with optical jets (3C 15, 3C 66B, 3C 78, and 3C 264) exceptionally reveal the presence of circular, face-on dust discs, which allow us to determine jet inclination as being very close to the line of sight. Similar FR I sources without optical jets do not reveal such a dust structures. Scarpa and Urry (\cite{sca02}) analyzed energy budgets of optical jet sources and found that they are more strongly beamed than the average population of extragalactic radio sources, with the average jet bulk Lorentz factor $\Gamma \sim 7.5$ and the mean jet viewing angle $\theta \sim 20^0$. Similar conclusions were presented by Jester (\cite{jes03}), who underlined that optical jets seem to be common between all types of radio loud AGNs, and that the small number of the detected ones is solely due to surface brightness selection effects connected with relativistic beaming. All these conclusions are strongly and directly supported by the {\it HST} observations of the apparent superluminal motions at the kpc scale in the M 87 jet (Biretta et al.~\cite{bire99}).

As shown by Heinz and Begelman (\cite{hei97}) the bulk Lorentz factor of the M 87 jet $\Gamma \sim 2 - 3$ coupled with a sub-equipartition magnetic field, $B \sim 0.2 - 0.7 \, B_{eq}$, and adiabatic fluctuations of the jet plasma can easily explain spectral behavior of the optical synchrotron emission along the jet flow, and in particular lack of the observed synchrotron cooling signatures. Thus, as compared to conclusions of Meisenheimer et al. (\cite{mei96}) there remains a general picture in which the magnetic field controls both emissivity and spectral index changes, although the requirement for particle re-acceleration vanishes. Continuing this approach, Scarpa and Urry (\cite{sca02}) claimed that highly relativistic bulk velocities, inferred by them for large-scale optical jets, can solve \emph{on average} the electron lifetime problem without need for particle re-acceleration, even if the equipartition condition is fulfilled. That is because relativistic bulk velocities blue-shifting photon energies in the observer rest frame, $\nu = \delta \, \nu'$, and amplifying the observed synchrotron luminosity, $L = (\delta^3 / \Gamma) \, L'$, decrease both the required energy of the electrons responsible for the optical synchrotron emission as well as an equipartition value for the magnetic field intensity as compared to the one computed for nonrelativistic bulk velocities, $B_{{\rm eq}} = B_{{\rm eq}, \, \delta=1} \, \delta^{-5/7}$ (see Stawarz et al.~\cite{sta03}). Hence, the rate of synchrotron cooling is also decreased, and the deprojected lengths of the optical jets, $l / \sin \theta$, is consistent \emph{on average} with the propagation length of the optically emitting electrons, $l_{{\rm O}} \sim c \, \Gamma \, t_{{\rm life, \, O}}'$. However, as pointed by Jester et al. (\cite{jes01}), this is not the case for the 3C 273. It is due to the fact, that with the increasing jet bulk Lorentz factor the energy density of the CMB radiation as measured in the jet rest frame also increases, $U_{{\rm CMB}} = \Gamma^2 \, U_{{\rm CMB}}$, leading to efficient inverse-Compton losses, and hence to the appropriate propagation length (for small jet viewing angles and nedlected redshift corrections) $l_{{\rm O}} \sim 0.3 \, \Gamma^{3/2} \, B_{-4}^{-3/2} \, \nu_{15}^{-1/2} \, ( 1 + 10^{-3} \, \Gamma^2 \, B_{-4}^{-2})^{-1}$ kpc. As a result, no combination of the $B < B_{{\rm eq}}$ and $\delta > 1$ can resolved the electron lifetime problem within the long (more than ten times longer than in M 87) jet in 3C 273. Thus, some kind of particle re-acceleration acting continuously along the jet body seems to be still required, even if the optical large-scale jets are highly relativistic and/or the jet magnetic field is below the equipartition value.

\section{X-ray emission}
\label{sect:Xra}
Before {\it CXO} mission, only four extragalactic large-scale X-ray jets were known. The very first object of this kind was discovered in the closest FR I radio galaxy Centaurus A even before detection of its radio counterpart (Schreier et al.~\cite{sch79}, Feigelson et al.~\cite{fei81}). The two other ones were, not surprisingly, the M 87 jet (Schreier et al.~\cite{sch82}, Biretta et al.~\cite{bire91}, Neumann et al.~\cite{neu97}) and the 3C 273 one (Harris and Stern~\cite{har87}, R\"oser et al.~\cite{ros00}). Finally, X-ray emission of the knot in the large-scale jet of peculiar Seyfert galaxy 3C 120 was reported by Harris et al. (\cite{har99}). In all these cases the observed X-ray emission was recognized as a non-thermal one, being directly connected with the radio-emitting plasma activity. Further detections of the X-ray jets were however considered unlikely, because production of the nonthermal (synchrotron or inverse-Compton) keV photons by large-scale jets was supposed to be much lower than production of thermal X-ray radiation due to the hot gas surrounding extragalactic radio sources, except maybe the few closest and most powerful objects. Therefore, the first {\it CXO} discovery of the extremely extended (projected length $> 100$\,kpc) X-ray jet associated with the distant ($z=0.654$) quasar PKS 0637-752 was a surprise (Chartas et al.~\cite{cha00}, Schwartz et al.~\cite{sch00}). Since then, {\it CXO} detected about 30 such objects in different types of radio-loud AGNs.

A lot of X-ray jets are observed by {\it CXO} in powerful quasars. They are the jets associated with 3C 273 (Marshall et al.~\cite{mar01}, Sambruna et al.~\cite{sam01}), PKS 0637-752 (Chartas et al.~\cite{cha00}, Schwartz et al.~\cite{sch00}), 3C 207 (Brunetti et al.~\cite{bru02a}), 3C 9 (Fabian et al.~\cite{fab03}), Q0957+561 (Chartas et al.~\cite{cha02}), PKS 1136-135, PKS 1150+497, PKS 1354+195, and 3C 179 (Sambruna et al.~\cite{sam02}), PKS 0605-085, PKS 1510-089, 3C 345, PKS 1642+690, PKS 1928+738 (Sambruna et al.~\cite{sam04b}), including two GPS sources PKS 1127-145 and B2 0738+313 (Siemiginowska et al.~\cite{sie02},~\cite{sie03a}, respectively). There are also the X-ray jets in the FR II radio galaxies Pictor A (Wilson et al.~\cite{wil01}), 3C 303 (Kataoka et al.~\cite{kat03a}) and 3C 371 (Pesce et al.~\cite{pes01}), in the FR II/FR I sources 3C 15 (Kataoka et al.~\cite{kat03b}) and NGC 6251 (Sambruna et al.~\cite{sam04a}), in PKS 0521-365 (Birkinshaw et al.~\cite{birk02}) and in the Seyfert radio galaxy 3C 120 (Harris et al.~\cite{har99}). The X-ray jets are also common between less-luminous FR I systems like M 87 (Marshall et al.~\cite{mar02}, Wilson and Yang~\cite{wil02}), Centaurus A (Kraft et al.~\cite{kra00},~\cite{kra02}, Hardcastle et al.~\cite{hard03}), 3C 31 (Hardcastle et al.~\cite{hard02}), B2 0206+35 and B2 0755+37 (Worrall et al.~\cite{wor01}), 3C 66B (Hardcastle et al.~\cite{hard01}), 3C 129 (Harris et al.~\cite{har02a}), M 84 (Harris et al.~\cite{har02b}), 3C 270 (Chiaberge et al.~\cite{chi03}), NGC 315 (Worrall et al.~\cite{wor03}) and Seyfert galaxy 3C 120 (Harris et al.~\cite{har99}). All these jets differ significantly in morphology and spectral properties, but as mentioned above, their featureless X-ray emission is uniquely considered as a manifestation of the non-thermal radiation processes. The thermal bremsstrahlung models can be excluded, as they require the existence of clumps of over-pressured gas far from the active galactic centers, with no obvious source of confinement and with the gas densities exceeding the ones implied by Faraday rotation and depolarization measurements (e.g., Harris and Krawczynski~\cite{hk02}). However, the nature of nonthermal processes responsible for production of relatively strong X-ray emission in the considered objects is still under debate. One can expect, that different processes can play a role in cases of low-power and high-power sources, as observed properties of the X-ray jets are much different in these two cases.

X-ray jets in FR I radio sources are usually quite short (projected lengths $\sim 1 - 5$\,kpc), being composed from diffusive knots with detectable inter-knot emission. X-ray jet morphology corresponds roughly to the radio morphology, and, if observed, to the optical one (M 87, PKS 0521-315, 3C 66B, 3C 31 and B2 0755+37). A very interesting and important difference between X-ray and radio/optical maps is the position of some knot maxima as observed at these frequencies, with the X-ray peaks placed closer to the jet base. The observed spatial offsets are $\sim 0.008$\,kpc in M 87, $\leq 0.08$\,kpc in Centaurus A and $\sim 0.2$\,kpc in 3C 66B. Except the weakest and the smallest objects (3C 270 and M 84), the observed X-ray luminosities of the discussed jets are $L_{\rm X} \sim 10^{39} - 10^{42}$\,erg s$^{-1}$ while the X-ray spectral index is on average $\alpha_{\rm X} \sim 1.3$, ranging from $\sim 1.1$ (in 3C 31) to $\geq 1.5$ (e.g., NGC 315, some knots in M 87 and Centaurus A). This suggests that radio-to-X-ray emission is consistent with a broken power-law continuum, where the low-energy part extends from radio up to some break frequency $\nu_{{\rm br}}$ placed near optical frequencies (Sect. 3) with a power-law slope $\alpha_{\rm 1} < 1$, and then it continues up to at least X-ray frequencies with $\alpha_{\rm 2} > 1$. This in turn suggests that X-ray photons in FR I jets are synchrotron in nature. One should note, that the observed values of the spectral break are larger than the expected ones in a simple cooling scenario, $\Delta \alpha = \alpha_{\rm 2} - \alpha_{\rm 1} = 0.5$.

The synchrotron nature of the FR I jets X-ray emissions implies presence of the highly relativistic electrons, with the Lorentz factor $\gamma > 10^7$ for the equipartition magnetic field $10^{-4}$\,G, and hence with the propagation length $l_{{\rm X}} \leq 0.03$\,kpc, much shorter than the projected lengths of the observed X-ray diffusive structures. This fact substantially strengthens a requirement of particle (re-)acceleration acting continuously along the jet, even if bulk jet velocities are relativistic and the jet magnetic fields are below equipartition. An alternative possibility would be the production of the X-ray photons via inverse-Compton processes. However, there are a few reasons why it is unlikely. First, the inverse-Compton models applied to the considered sources usually underestimate the observed X-ray fluxes, unless extreme jet parameters are invoked (e.g. Hardcastle et al.~\cite{hard01}; cf. Celotti et al.~\cite{cel01}). Secondly, as pointed out by Harris et al. (\cite{har03}), rapid radiative cooling of the high energy electrons responsible for the X-ray synchrotron emission is more consistent with relatively short variability time scale observed in the M 87 X-ray knots, which is of the order of years, than a much longer radiative losses time scale of low energy electrons in the inverse-Compton scenario (see also Harris et al.~\cite{har97}). Finally, also due to differences in the involved electron energies, the observed steep X-ray spectral indices -- expected in a framework of synchrotron emission of radiatively cooling electrons -- contradict the inverse-Compton models. To sum up, {\it CXO} observations of the large-scale FR I jets indicate that these objects can efficiently accelerate electrons up to at least $100$\,TeV energies.

Difficulties are encountered in attempts of modeling the FR I jet X-ray emission in a framework of the standard shock-in-knot scenario, beginning with the observed steep spectral indices $\alpha_{\rm X} > \alpha_{\rm R} + 0.5$ and the X-ray/radio (optical) offset problem. It was initially proposed that the latter effect could be explained by particle acceleration at the shock front, placed at the position of X-ray peak, and subsequent cooling of the radiating particles as they propagate downstream along the jet, where the radio emission reaches its maximum. Hence, differences in the positions of the X-ray/radio maxima would simply be caused by differences in cooling times of the electrons radiating at different frequencies (Hardcastle et al.~\cite{hard01}, Bai and Lee~\cite{bai03}). However, detailed {\it CXO} observations of the Centaurus A jet revealed that this picture of a uniform downstream particle advection is not adequate (Hardcastle et al.~\cite{hard03}). A possible solution for the problem might be a spatially non-uniform magnetic field, increasing within some knot regions, together with efficient stochastic particle acceleration acting thereby to generate particles with different energies along the flow. Interestingly, such magnetic field inhomogeneous structures could also cause the observed large values of spectral breaks between low- and high-energy part of the observed synchrotron spectra, $\Delta \alpha > 0.5$ (Cavallo et al.~\cite{cav80}, Coleman and Bicknell~\cite{col88}). If this picture is correct, the X-ray emission -- its spatial distribution and spectral properties -- follows directly inhomogeneous magnetic field structures within the FR I jets, and also possibly within more powerful sources.

In the hybrid morphology radio galaxy PKS 0521-365 the X-ray jet is placed on the source FR I side, and its spectral properties resemble those seen in the FR I jets: the steep spectral index $\alpha_{{\rm X}} \sim 1.3$, the X-ray luminosity $L_{\rm X} \sim 10^{42}$\,erg s$^{-1}$, and the projected length $\sim 3$\,kpc. Also, the X-ray jet in the intermediate FR II/FR I source 3C 15 is not much different from the ones observed in FR I objects, however with a smaller spectral break implied by indices $\alpha_{{\rm R-O}} \sim 0.9$ and $\alpha_{{\rm O-X}} \sim 1.1. - 1.2$, as well as with a flatter X-ray spectral index $\alpha_{\rm X} \sim 0.7$. In the FR II radio galaxy with a BL Lac nucleus 3C 371, the situation is similar. Here the X-ray jet is also a few kpc long, with edge-darkened X-ray morphology -- opposite to the radio one -- and with the X-ray flux below the extrapolated radio-to-optical continuum. In the classical double radio source Pictor A, the projected X-ray jet is more than $100$\,kpc long, only twice shorter than its weak radio counterpart. The X-ray emission of Pictor A jet is smooth (continuous), with the X-ray spectral index, $\alpha_{{\rm X}} \sim 0.94$, similar to the radio spectral index, $\alpha_{\rm R} \sim 0.85$, and also to the radio-to-X-ray power-law slope, $\alpha_{{\rm R-X}} \sim 0.87$ (optical upper limits are consistent with one power-law connecting radio and X-ray fluxes). Therefore, in all large-scale jets observed by {\it CXO} in FR II and intermediate type radio sources, the X-ray jet emission seems to be consistent with the synchrotron model (the X-ray knots in 3C 303, located several kpc from the parent galaxy, are too faint to discuss their spectral properties). In general, these jets are on average longer, more luminous, exhibiting flatter X-ray spectra and smaller spectral breaks than the earlier described FR I jets. For the recently discovered X-ray knot in FR II/ FR I radio galaxy NGC 6251 the synchrotron model is also the most plausible option, although in such a case spectral hardenings at the highest electron energies are required, as the X-ray continuum is extremely flat (X-ray spectral index $\alpha_{\rm X} \sim 0.06^{+0.43}_{-0.38}$).

Large-scale X-ray jets in powerful quasars differ significantly from the respective structures observed in FR Is. First, they are long, with projected lengths or distances of the knots from the host galaxies ranging from few tens of kpcs up to more than $100$\,kpc (in PKS 0637-752, PKS 1127-145, B2 0738+313). In general, their knotty morphology corresponds well to the radio and eventually optical maps, although for some sources important differences were noted. The X-ray jets in quasars are also relatively luminous, with the $1 - 10$\,keV luminosity being on average $L_{\rm X} \sim 10^{43} - 10^{45}$\,erg s$^{-1}$. Their X-ray spectra are flat (sometimes very flat, possibly even inverted), with the spectral indices ranging from $\alpha_{\rm X} \sim 0.1$ in 3C 207, PKS 0605-085 or PKS 1510-089 up to $\sim 0.85$ in PKS 0637-752\footnote{In the case of PKS 1928+738 the X-ray spectral index is exeptionally high, $\alpha_{\rm X} \sim 1.66$.}. One of the most characteristic feature of the broad-band spectral energy distribution in the discussed objects is the X-ray flux placed usually (although not always) above the extrapolated radio-to-optical continuum. One should also mention, that in a few cases (e.g., 3C 273, Q0957+561, PKS 1127-145) the observed X-ray flux decreases along the jet, opposite to the radio emission. Also, in the GPS quasars PKS 1127-145 and B2 0738+313, as well as in 3C 273, significant offsets between positions of the X-ray and radio knots were noted, similar but much larger ($> 10$\,kpc for PKS 1127-145) than the ones observed in some FR I jets. Finally, one should emphasize that knots observed in the discussed objects at radio, optical and X-ray photon energies are of similar sizes ($\sim 1$\,kpc), and of similar flux profiles, i.e. no frequency-dependence in the knot extent was noted.

As discussed by Chartas et al. (\cite{cha00}) and Schwartz et al. (\cite{sch00}), in the case of PKS 0637-752 a single-power-law synchrotron spectrum connecting radio, optical and X-ray emission can be excluded, as the X-ray flux is above the extrapolated radio-to-optical continuum about two orders of magnitude. On the other hand, the inverse-Compton scattering of the host galaxy stellar emission, of the AGN radiation or -- for nonrelativistic jet velocities -- of the CMB photons significantly underestimates the observed X-ray luminosity. The synchrotron-self Compton emission can possibly match the required X-ray flux, but only if the jet magnetic field is far below the equipartition (with $\delta \sim 1$), or if the jet emission is de-beamed in a sense that $\delta < 1$. Yet another possibility is that X-ray photons are produced by a different electron population than the one responsible for radio-to-optical emission. Tavecchio et al. (\cite{tav00}) and Celotti et al. (\cite{cel01}) proposed instead, that the PKS 0637-752 jet spectrum can be explained if one assumes that the jet is highly relativistic on hundreds-of-kpc scale, with the bulk Lorenz factor $\Gamma > 10$ and the jet viewing angle $\theta < 6^0$, similar to what is observed at parsec scales in this source. For such jet parameters, the low-energy electrons (if present) can efficiently inverse-Compton upscatter CMB photons to the observed frequency $\nu_{{\rm ic}} \sim \delta^2 \, \nu_{{\rm CMB}} \, \gamma^2$, resulting in relatively strong X-ray emission with the observed luminosity $L_{{\rm ic}} \sim (\delta/ \Gamma)^2 \, (U'_{{\rm CMB}} / U'_B) \, L_{{\rm syn}}$, where $U'_{{\rm CMB}} = 4 \cdot 10^{-13} \, \Gamma^2 \, (1+z)^4$\,erg cm$^3$ is the energy density of the CMB radiation as measured in the jet rest frame at the redshift $z$. When applied to PKS 0637-752 jet, this model is consistent with the observed flat spectral index of the X-ray emission, $\alpha_{\rm X} \sim \alpha_{\rm R}$, and is favorable on the grounds of the energetic arguments, as it allows for magnetic field close to equipartition and implies a much lower jet power as compared to the de-beamed SSC model (see also Ghisellini and Celotti~\cite{ghi01}). Therefore, comptonization of the CMB radiation within highly relativistic large-scale quasar jets was widely discussed in the literature, being applied to all quasar jets observed by {\it CXO} (see also Harris and Krawczynski~\cite{hk02}, Sahayanathan et al.~\cite{sah03}).

One of the predictions of the model explaining quasar jets X-ray emission as inverse-Compton scattering of CMB photons is that the X-ray quasar jets should be detectable by {\it CXO} with the same surface brightness anywhere at high redshifts (Schwartz~\cite{sch02a}). This is because the increase of the CMB energy density $\propto(1+z)^4$ exactly balances the decrease of the surface brightness of the extended jet structure due to expansion of the Universe, $\propto (1+z)^{-4}$. Thus, a sample of similar (in luminosity and magnetic field intensity) quasar jets placed over a range of redshift should exhibit dependence $L_{\rm R} / L_{\rm X} \propto (1+z)^{-4}$. It is interesting in this context that the large-scale X-ray jets seem to be common among distant quasar sources, with the X-ray luminosity being usually a few percent of the core luminosity (Schwartz et al.~\cite{sch03}). The most distant ones are found currently at the redshifts of $z=2.012$ (3C 9) and $z = 4.3$ (GB 1508+5714, Siemiginowska et al.~\cite{sie03b}). There is also an unconfirmed detection of the X-ray jet associated with one of the three most distant quasars known, SDSS 1030+0524 at $z=5.99$ (Schwartz~\cite{sch02b}). In the last case there is no detection of the radio jet, while for the quasar GB 1508+5714 the jet radio structure was discovered after the X-ray detection (Cheung~\cite{che04}). Interestingly, it indicates that X-rays can be more useful in studying extremely distant large-scale jets than the radio waves.

Now, let us briefly mention some `warnings' to the model discussed above. First, it assumes highly relativistic velocities on large scales, which contradict the radio observations (Sect. 2), but on the other hand is supported by the optical results (Sect. 3). As mentioned above, the discrepancy can be explained by assuming the velocity structure across the jet flows. Yet, such high bulk velocities require very small jet inclinations, unless extreme intrinsic luminosities are invoked in order to balance the Doppler-hiding of the jet emission. This requirement, in turn, implies that the deprojected lengths of the considered sources have to be in some cases as large as one megaparsec (PKS 0637-752, PKS 1127-145, B2 0738+313). Such large linear sizes are not common in quasars or FR II objects, except a class of weak extended radio structures observed in the so-called Giant Radio Sources. Also, small jet viewing angles are not expected in the cases of lobe dominated quasars like 3C 207 and 3C 9. Secondly, in a framework of the discussed model, the X-ray emission is produced by a low-energy part of the electron energy distribution, with the Lorentz factors $\gamma < 10^3$, whose radio synchrotron emission cannot be directly observed. In addition, this unknown low-energy part of the electron spectrum has to cut-off sharply at $\gamma \sim 10$ in order not to overproduce optical radiation by interactions with the CMB photons.

Long radiative cooling time scale of $\gamma < 10^3$ electrons is consistent with large extents of the X-ray quasar jets. At the same time it also leads to another important problem, because knotty morphologies of the considered sources contradict long cooling times of radiating electrons. As presented above, the X-ray knot extensions are similar to the radio and optical ones, while the propagation lengths of the electrons responsible for emission at these photon energy ranges are much different. As discussed by Tavecchio et al. (\cite{tav03}), this cannot be explained solely by invoking adiabatic losses, if the knots are in fact homogeneous steady regions of particle acceleration, as usually modeled. This led Tavecchio et al. to suggest that knots are formed by several smaller `clumps', which are themselves strongly matter dominated, and which can adiabaticaly expand rapidly enough to account for the observed knot morphology. Note however, that in a framework of the SSC emission, by postulating inhomogeneous structure of the knot regions and departures from the energy equipartition one can try to explain {\it CXO} quasar jet observations. Stawarz et al. (\cite{sta04}) proposed instead, that high knot-to-interknot brightness contrasts, frequency-independent knot profiles and almost universal extents of the knot regions in the large-scale quasar jets observed at radio, optical and X-rays, can be explained (independently of the exact emission process producing X-ray photons) if the knots represent relativistically moving portions of the jet, with an excess kinetic power due to intermittent activity of the jet engine. Stawarz et al. proposed the appropriate time scales for jet activity and quiescent epochs $\sim 10^4$ yr and $\sim 10^5$ yr, respectively (see in this context Reynolds and Begelman~\cite{rey97}, and also Marecki et al.~\cite{mar03}).

Obviously, there is still a place to look for other mechanisms that might be able to produce non-thermal X-ray photons within the large-scale quasar jets, and for the ways to discriminate between different models. In particular, detection of the strong X-ray counterjet to any of the considered objects would exclude significant beaming of the observed emission. Up to now, X-ray counterjets are possibly found in weak FR I source 3C 270, as well as in distant, lobe dominated quasar 3C 9. In the latter case, it is unclear whether the X-ray emission is non-thermal in nature, as -- apart from the low photon statistic preventing detailed spectral analysis -- there is no one-to-one morphological correspondence between the X-ray and radio structures. Similarly, in the X-ray maps of classical double Cygnus A system, linear features aligned with the radio jet and the counterjet were noted, although their direct connection with the radio jet plasma is unclear (Smith et al.~\cite{smi02}). It is also possible that in both 3C 9 and Cygnus A the aligned X-ray structures are due to thermal radiation of shocked gas which confines the jet plasma, as was suggested by Carilli et al. (\cite{car02}) for the case of PKS 1138-262. Let us note, that such diffusive X-ray emission, as well as X-ray flux above the extrapolated radio-to-optical continuum is expected in a framework of the model discussed by Stawarz and Ostrowski (\cite{sta02b}).

Stawarz and Ostrowski (\cite{sta02a,sta02b}) proposed the model, in which electrons trapped within highly turbulent shear boundary layers of the large-scale jets are piled-up near the maximum energy $\gamma \sim 10^8$, producing the flat synchrotron X-ray spectrum (see also Ostrowski~\cite{ost00}). This idea received significant support recently from detailed {\it CXO} observations of the Centaurus A, which revealed a clear edge-brightening structure of the X-ray jet at kpc scale (Hardcastle et al.~\cite{hard03}), and also from detailed {\it HST} analysis of 3C 273 jet by Jester et al. (\cite{jes03}), where spectral hardening of the jet synchrotron spectrum starting at UV frequencies was measured. The latter observation suggests smooth continuation of the synchrotron optical emission up to higher photon energies not in terms of single (or broken) power-law, but involving instead spectral hardenings at high frequencies (see in this context Dermer and Atoyan~\cite{der02}). Also, extremely flat (possibly even inverted) X-ray spectral indices found in a few quasar jets (Sambruna et al.~\cite{sam04b}), which make a problem for shock acceleration, are expected to occur in a framework of a boundary layer acceleration scenario. In addition, if the X-rays from these jets are synchrotron in nature (what most probably is the case in 3C 120, and in the brightest knot of 3C 273), one has to invoke extended particle acceleration sites -- and not the localized shock fronts -- in order to overcome the problem of flat X-ray spectral indices observed in face of strong cooling of the X-ray emitting electrons (see discussion in Aharonian~\cite{aha02}). This problem could be also explained by postulating multiple shock accelerations within knot regions (see Stawarz et al.~\cite{sta04}). Yet another possibility can be synchrotron emission of extremely energetic protons, with energies $\geq 10^{18}$\,eV, as discussed by Aharonian (\cite{aha02}). This model can possibly explain long extensions of the X-ray quasar jets, as the synchrotron cooling time for such protons in the required magnetic field $\sim 10^{-3}$ G\,is very long, $10^7 - 10^8$\,yr. Let us finally note, that the other kinds of hadronic models, like proton induced cascades discussed, e.g., by Mannheim et al. (\cite{man91}), can hardly explain {\it CXO} data (Aharonian~\cite{aha02}).

\section{$\gamma$-ray emission}
\label{sect:Gam}
The possibility of observing extragalactic large-scale jets at $\gamma$-ray frequencies was not extensively discussed in the literature, because the presence of very high-energy particles needed to produce $\gamma$-rays within the considered objects at the level allowing for positive detection, was considered unlikely. Recent optical and X-ray observations, as well as development of modern ground-based and space $\gamma$-ray telescopes, can permit the study of large-scale jets as high energy $\gamma$-ray sources. Even non-detections resulting in flux upper limits are useful for discrimination between different models proposed in the literature to explain radiation of these hardly known objects at lower energies. However, the primary difficulty in discussing $\gamma$-ray radiative output of the large-scale jets is that even though present and future $\gamma$-ray telescopes are sensitive enough to eventually detect the $\gamma$-ray signal from the discussed objects, their angular resolution will not allow for the separation of the kpc-scale jet emission from the possible core component.

Until now, the possible $\gamma$-ray detections of the large-scale jets are connected with two nearby FR I objects, Centaurus A and M 87. {\it CANGAROO} observations of the Centaurus A system put the upper limit on its emission at the very high energy (VHE) $\gamma$-ray band, $S(\varepsilon_{\gamma} \geq 1.5 \, {\rm TeV}) < 1.28 \cdot 10^{-11}$\,ph cm$^{-2}$ s$^{-1}$ (Rowell et al.~\cite{row99}). However, the discussed object was detected in the past at TeV energies, with the observed flux $S(\varepsilon_{\gamma} \geq 0.3 \, {\rm TeV}) \sim 4.4 \cdot 10^{-11}$\,ph cm$^{-2}$ s$^{-1}$ (Grindlay et al.~\cite{gri75}), what made Centaurus A the very first (although not confirmed) detected extragalactic source of VHE radiation. The recent upper limit suggests that this emission can be variable on a timescale of years, which in turn may suggest that $\gamma$-rays are produced at small scales, close to the active nucleus. The core origin of the detected $\gamma$-rays can be justified by studying the spectral energy distribution of the broad-band core emission, although not without some controversies. For example, detailed analysis of the Centaurus A central region emission by Chiaberge et al. (\cite{chi01}) indicates that in the framework of the leptonic model the small-scale Centaurus A jet is not expected to be the source of VHE photons, while Bai and Lee (\cite{bai01}) have drawn an opposite conclusion. Such controversy is also present in the case of M 87 radio galaxy, for which recent {\it HEGRA} observations resulted in a positive detection of the VHE emission with the observed flux $S(\varepsilon_{\gamma} \geq 0.73 \, {\rm TeV}) \sim 0.96 \cdot 10^{-12}$\,ph cm$^{-2}$ s$^{-1}$ (Aharonian et al.~\cite{aha03}). This emission was proposed to originate from hadronic processes in either small-scale jet (Reimer et al.~\cite{rei04}) or interstellar medium (Pfrommer and Ensslin~\cite{pfr03}). However, one should note that the large-scale jets in both M 87 and Centaurus A are natural sites for production of the VHE $\gamma$-ray photons.

Stawarz et al. (\cite{sta03}) discussed the VHE $\gamma$-ray emission of the low-luminosity kpc-scale jets observed by {\it HST} and {\it CXO} in FR I radio galaxies. These objects are most probably relativistic, and no doubt able to produce very high-energy electrons. The energy distribution of such electrons is relatively well covered by the radio-to-X-ray observations (Sects. 2 - 4), and therefore one can reasonably estimate the expected VHE $\gamma$-ray fluxes resulting from the inverse-Compton processes. For the Centaurus A kpc-scale jet, Stawarz et al. (\cite{sta03}) estimated measurable by future $\gamma$-ray telescopes emission at $1 - 10$\,GeV and $0.1 - 1$\,TeV photon energies due to comptonization of the hidden blazar radiation and the synchrotron self-Compton process, respectively. Comptonisation of the blazar radiation within large-scale jets in radio galaxies was discussed also by Celotti et al. (\cite{cel01}) in the context of their X-ray radiation. A new target photon field for the inverse-Compton process within the kpc-scale jet was considered by Stawarz et al. (\cite{sta03}), who showed that the VHE emission detected recently from the M 87 system can result from comptonization of stellar and circumstellar infrared photons of the host galaxy.

The analysis by Stawarz et al. (\cite{sta03}) suggests that a variety of inverse-Compton processes can control spectral properties of the VHE emission in different FR I objects. Obviously, one should expect significant differences between the FR I and quasar jets, with different radio-to-X-ray spectra suggesting different $\gamma$-ray radiative outputs. In addition, as quasars are cosmologically distant objects, the effects of absorption of their $\gamma$-ray emission on cosmic background microwave-to-optical radiation fields can play an important role in controlling the observed $\gamma$-ray spectra (see in this context Aharonian et al.~\cite{aha94}).

\section{Conclusions}
\label{sect:Con}
Extragalactic large-scale jets are confirmed sources of radio, optical and X-ray photons. Recent optical and X-ray observations have already substantially enriched our knowledge of the discussed objects, and one should expect many new and important results to come in the near future. The discovery of the X-ray jets without radio counterparts, as well as prospects of observing extragalactic jets at $\gamma$-ray frequencies, are particularly exciting in this respect.

Thanks to {\it HST} and {\it CXO} we collected convincing evidence that extragalactic jets remain relativistic at tens and hundreds of kiloparsec scales. We know that they possess internal velocity profiles and magnetic field inhomogeneous structures, which seem to be closely related to the energy dissipation processes. Particle acceleration processes acting in jets are much less understood than previously thought. However, we take for certain that particle re-acceleration within the large-scale jets is inevitable. In particular, optical and X-ray observations suggest that such processes are more complex than usually modeled, and it is possible that there are many different ways of dissipating the jet kinetic energy to the radiating particles to be considered. For example, except for the localized particle acceleration sites within the knot regions, particle acceleration processes seem to act continuously along the whole jet body, efficiently producing very high-energy electrons. In addition, a variety of spectral shapes observed in different types of extragalactic large-scale jets indicates that the energy distribution of the accelerated particles does not follow a simple power-law behavior. Thus, spectral and morphological characteristics of the large-scale jets multiwavelength emission are extremely important for understanding the long-discussed issue of particle acceleration in astrophysical jets.

\begin{acknowledgements}
I am grateful to Micha\l \, Ostrowski and Marek Sikora for their help and collaboration, and to many colleagues with whom I had opportunity to discuss the jet phenomenon. The present work was supported by Komitet Bada\'{n} Naukowych through the grant PBZ-KBN-054/P03/2001.
\end{acknowledgements}


\begin{thebibliography}
\baselineskip=12pt
\parskip=-3pt

\bibitem[2002]{aha02} Aharonian F. A., 2002, \mnras, 332, 215
\bibitem[1994]{aha94} Aharonian F. A., Coppi P. S., V\"olk, H. J., 1994, \apj, 423, L5
\bibitem[2003]{aha03} Aharonian F. A. et al., 2003, \aap, 403, L1
\bibitem[1954]{baa54} Baade W., Minkowski R., 1954, \apj, 119, 215
\bibitem[1995]{bah95} Bahcall J. N., Kirhakos S., Schneider D. P., Davis R. J., Muxlow T. W. B., Garrington S. T., Conway R. G., Unwin S. C., 1995, \apj, 452, L91
\bibitem[2001]{bai01} Bai J. M., Lee M. G., 2001, \apj, 549, 173
\bibitem[2003]{bai03} Bai J. M., Lee M. G., 2003, \apj, 585, 113
\bibitem[1982]{beg82} Begelman M.C., 1982, In: D. S. Heeschen and C. M. Wade, eds., Proc. IAU Symp. 97, Extragalactic radio sources, p. 223
\bibitem[1984]{beg84} Begelman M. C., Blandford R. D., Rees M. J., 1984, Rev.Mod.Phys., 56, 255
\bibitem[1994]{bic94} Bicknell G. V., 1994, \apj, 422, 542
\bibitem[1982]{bic82} Bicknell G. V., Melrose D. B., 1982, \apj, 262, 511
\bibitem[1996]{bic96} Bicknell G. V., Begelman M. C., 1996, \apj, 467, 597
\bibitem[1991]{bire91} Biretta J. A., Stern C. P., Harris D. E., 1991, \aj, 101, 1632
\bibitem[1999]{bire99} Biretta J. A., Sparks W. B., Macchetto F., 1999, \apj, 520, 621
\bibitem[2000]{bir00} Birk G. T., Lesch H., 2000, \apj, 530, L77
\bibitem[1991]{birk91} Birkinshaw M., 1991, \mnras, 252, 505
\bibitem[2002]{birk02} Birkinshaw M., Worrall D. M., Hardcastle M. J., 2002, \mnras, 335, 142
\bibitem[1979]{bla79} Blandford R., Konigl A., 1979, \apj, 232, 34
\bibitem[1987]{bla87} Blandford R., Eichler D., 1987, Phys.Rev., 154, 1
\bibitem[2001]{blu01} Blundell K. M., Rawlings S., 2001, \apj, 562, L5
\bibitem[1984]{bri84} Bridle A. H., Perley R. A., 1984, ARA\&A, 22, 319
\bibitem[2002]{bru02b} Brunetti G., 2002, preprint (astro-ph/0207671)
\bibitem[2002]{bru02a} Brunetti G., Bondi M., Comastri A., Setti G., 2002, \aap, 381, 795
\bibitem[1980]{but80} Butcher H. R., van Breugel W., Miley G. K., 1980, \apj, 235, 749
\bibitem[1996]{car96} Carilli C. L., Barthel P. D., 1996, A\&ARv, 7, 1
\bibitem[2002]{car02} Carilli C. L., Harris D. E., Pentericci L., Rottgering H. J. A., Miley G. K., Kurk J. D., van Breugel W., 2002, \apj, 567, 781
\bibitem[1980]{cav80} Cavallo G., Horstman H. M., Muracchini A., 1980, \aap, 86, 36
\bibitem[2001]{cel01} Celotti A., Ghisellini G., Chiaberge M., 2001, \mnras, 321, L1
\bibitem[2000]{cha00} Chartas G., Worrall D. M., Birkinshaw M., Creistello-Dittmar M., Cui W., Ghosh K. K., Harris D. E., Hooper D. J., Jauncey D. L., Kim D. W., Lovell J., Marshall H. L., Mathur S., Schwartz D. A., Tingay S. J., Virani S. N., Wilkes B. J., 2000, \apj, 542, 655
\bibitem[2002]{cha02} Chartas G., Gupta V., Garmire G., Jones C., Falco E. E., Shapiro I. I., Tavecchio F., 2002, \apj, 565, 96
\bibitem[2002]{che02} Cheung C. C., 2002, \apj, 581, L15
\bibitem[2004]{che04} Cheung C. C., 2004, \apj, 600, L23
\bibitem[2001]{chi01} Chiaberge M., Capetti A., Celotti A., 2001, \mnras, 324, 33
\bibitem[2003]{chi03} Chiaberge M., Gilli R., Macchetto F. D., Sparks W. B., Capetti A., 2003, \apj, 582, 645
\bibitem[1992]{cla92} Clarke D. A., Bridle A. H., Burns J. O., Perley R. A., Norman M. L., 1992, \apj, 285, 173
\bibitem[1988]{col88} Coleman C. S., Bicknell G. V., 1988, \mnras, 230, 497
\bibitem[1993]{cra93} Crane P., Peletier R., Baxter D., Sparks W. B., Albrecht R., Barbieri C., Blades J. C., Boksenberg A., Deharveng J. M., Disney M. J., Jakobsen P., Kamperman T. M., King I. R., Macchetto F., Mackay C. D., Paresce F., Weigelt G., Greenfield P., Jedrzejewski R., Nota A., 1993, \apj, 402, L37
\bibitem[2003]{cro03} Croston J. H., Birkinshaw M., Conway E., Davies R. L., 2003, \mnras, 339, 82
\bibitem[1918]{cur18} Curtis H. D., 1918, Lick Obs.Publ., 13, 31
\bibitem[1979]{dan79} Danziger I. J., Fosbury R. A. E., Goss W. M., Ekers R. D., 1979, \mnras, 188, 415
\bibitem[2002]{der02} Dermer C. D., Atoyan A. M., 2002, \apj, 568, 81
\bibitem[1997]{dev97} de Vries W. H., O'Dea C. P., Baum S. A., Sparks W. B., Biretta J., de Koff S., Golombek D., Lehnert M. D., Macchetto F., McCarthy P., Miley G. K., 1997, ApJS, 110, 191
\bibitem[1986]{dey86} De Young D. S., 1986, \apj, 307, 62
\bibitem[1993]{dey93} De Young D. S., 1993, \apj, 405, 13
\bibitem[1994]{dey94} Dey A., van Breugel W. J. M., 1994, \aj, 107, 1977
\bibitem[1979]{eil79} Eilek J. A., 1979, \apj, 230, 373
\bibitem[1982]{eil82} Eilek J. A., 1982, \apj, 254, 472
\bibitem[2002]{eil02} Eilek J. A., Hardee P. E., Markovic T., Ledlow M., Owen F. N., 2002, NewAR, 46, 327
\bibitem[2003]{fab03} Fabian A. C., Celotti A., Johnstone R. M., 2003 \mnras, 338, 7
\bibitem[1974]{fan74} Fanaroff B. L., Riley J. M., 1974, \mnras, 167, 31
\bibitem[1981]{fei81} Feigelson E. D., Schreier E. J., Delvaille J. P., Giacconi R., Grindlay J. E., Lightman A. P., 1981 \apj, 251, 31
\bibitem[1992]{fra92} Fraix-Burnet D., 1992, \aap, 259, 445
\bibitem[1997]{fra97} Fraix-Burnet D., 1997, \mnras, 284, 911
\bibitem[1999]{gab99} Gabuzda D. C., 1999, In: M. Ostrowski and R. Schlickeiser, eds., Plasma Turbulence and Energetic Particles in Astrophysics, p. 301
\bibitem[2001]{ghi01} Ghisellini G., Celotti A., 2001, \mnras, 327, 739
\bibitem[2001]{gio01} Giovannini G., Cotton W. D., Feretti L., Lara L., Venturi T., 2001, \apj, 552, 508
\bibitem[2000]{gop00} Gopal-Krishna, Wiita P. J., 2000, \aap, 363, 507
\bibitem[1975]{gri75} Grindlay J. E., Helmken H. F., Brown R. H., Davis J., Allen L. R., 1975, \apj, 197, 9
\bibitem[2001]{hard01} Hardcastle M. J., Birkinshaw M., Worrall D. M., 2001, \mnras, 326, 1499
\bibitem[2002]{hard02} Hardcastle M. J., Worrall D. M., Birkinshaw M., Laing R. A., Bridle A. H., 2002, \mnras, 334, 182
\bibitem[2003]{hard03} Hardcastle M. J., Worrall D. M., Kraft R. P., Forman W. R., Jones C., Murray S. S., 2003, \apj, 593, 169
\bibitem[1987]{har87} Harris D. E., Stern C. P., 1987, \apj, 313, 136
\bibitem[1997]{har97} Harris D. E., Biretta J. A., Junor W., 1997, \mnras, 284, 21
\bibitem[1999]{har99} Harris D. E., Hjorth J., Sadun A. C., Silverman J. D., Vestergaard M., 1999, \apj, 518, 213
\bibitem[2002]{hk02}  Harris D. E., Krawczynski H., 2002, \apj, 565, 244
\bibitem[2002a]{har02a} Harris D. E., Krawczynski H., Taylor G. B., 2002a, \apj, 578, 60
\bibitem[2002b]{har02b} Harris D. E., Finoguenov A., Bridle A. H., Hardcastle M. J., Laing R. A., 2002b, \apj, 580, 110
\bibitem[2003]{har03} Harris D. E., Biretta J. A., Junor W., Perlman E. S., Sparks W. B., Wilson A. S., 2003, \apj, 586, 41
\bibitem[1997]{hei97} Heinz S., Begelman M. C., 1997, \apj, 490, 653
\bibitem[1995]{hjo95} Hjorth J., Vestergaard M., Sorensen A. N., Grundahl F., 1995, \apj, 452, L17
\bibitem[1991]{hug91} Hughes P. A., ed., 1991, Beams and Jets in Astrophysics, Cambridge University Press
\bibitem[2003]{jes03} Jester S., 2003, NewAR, 47, 427
\bibitem[2001]{jes01} Jester S., R\"oser H.-J., Meisenheimer K., Perley R., Conway R., 2001, \aap, 373, 447
\bibitem[2002]{jes02} Jester S., R\"oser H.-J., Meisenheimer K., Perley R., 2002, \aap, 385, L27
\bibitem[2000]{kai00} Kaiser C. R., Schoenmakers A. P., R\"ottgering H. J. A., 2000, \mnras, 315, 381
\bibitem[2003a]{kat03a} Kataoka J., Edwards P., Georganopoulos M., Takahara F., Wagner S., 2003a, \aap, 399, 91
\bibitem[2003b]{kat03b} Kataoka J., Leahy J. P., Edwards P. G., Kino M., Takahara F., Serino Y., Kawai N., Martel N., 2003b, \aap, 410, 833
\bibitem[1988]{kee88} Keel W. C., 1988, \apj, 329, 532
\bibitem[1990]{kom90} Komissarov S. S., 1990, SvAL, 16, 284
\bibitem[1994]{kom94} Komissarov S. S., 1994, \mnras, 269, 394
\bibitem[2000]{kra00} Kraft R. P., Forman W. R., Jones C., Kenter A., Murray S. S., Aldcroft T., Elvis M., Evans I., Fabbiano G., Isobe T., Jerius D., Karovska M., Kim D.-W., Prestwich A., Primini F., Schwartz D., 2000, \apj, 531, L9
\bibitem[2002]{kra02} Kraft R. P., Forman W. R., Jones C., Murray S. S., Hardcastle M. J., Worrall D. M., 2002, \apj, 569, 54
\bibitem[1989]{kun89} Kundt W., 1989, In: K. Meisenheimer and H.-J. R\"oser, eds., Hot spots in extragalactic radio sources, p. 179
\bibitem[1977]{lac77} Lacombe C., 1977, \aap, 54, 1
\bibitem[1980]{lai80} Laing R. A., 1980, \mnras, 193, 439
\bibitem[1996]{lai96} Laing R. A., 1996, In: P. E. Hardee, A. H. Bridle and J. A. Zensus, eds., ASP Conf. Ser. 100, Energy transport in radio galaxies and quasars, p. 241
\bibitem[2002]{lai02} Laing R. A., Bridle A. H., 2002, \mnras, 336, 328
\bibitem[1999]{lai99} Laing R. A., Parma P., de Ruiter H. R., Fanti R., 1999, \mnras, 306, 513
\bibitem[1999]{lar99} Lara L., Feretti L., Giovannini G., Baum S., Cotton W. D., O'Dea C. P., Venturi T., 1999, \apj, 513, 197
\bibitem[1985]{lin85} Lind K. R., Blandford R. D., 1985, \apj, 295, 358
\bibitem[2002]{liu02} Liu F. K., Zhang Y. H., 2002, \aap, 381, 757
\bibitem[1991]{mac91} Macchetto F., Albrecht R., Barbieri C., Blades J. C., Boksenberg A., Crane P., Deharveng J. M., Disney M. J., Jakobsen P., Kamperman T. M., King I. R., Mackay C. D., Paresce F., Weigelt G., Baxter D., Greenfield P., Jedrzejewski R., Nota A., Sparks W. B., Miley G. K., 1991, \apj, 373, L55
\bibitem[1991]{man91} Mannheim K., Biermann P. L., Kruells W. M., 1991, \aap, 251, 723
\bibitem[2003]{mar03} Marecki A., Spencer R. E., Kunert M., 2003, PASA, 20, 46
\bibitem[2001]{mar01} Marshall H. L., Harris D. E., Grimes J. P., Drake J. J., Fruscione A., Juda M., Kraft R. P., Mathur S., Murray S. S., Ogle P. M., Pease D. O., Schwartz D. A., Siemiginowska A. L., Vrtilek S. D., Wargelin B. J., 2001, \apj, 549, L167
\bibitem[2002]{mar02} Marshall H. L., Miller B., Davis D., Perlman E., Wise M., Canizares C., Harris D. E., 2002, \apj, 564, 683
\bibitem[1998]{mar98} Martel A. R., Sparks W. B., Macchetto D., Biretta J. A., Baum S. A., Golombek D., McCarthy P. J., de Koff S., Miley G. K., 1998, \apj, 496, 203
\bibitem[1996]{mei96} Meisenheimer K., R\"oser H.-J., Schl\"otelburg M., 1996, \aap, 307, 61
\bibitem[1997]{mei97} Meisenheimer K., Yates M. G., R\"oser H.-J., 1997, \aap, 325, 57
\bibitem[1980]{mil80} Miley G. K., 1980, ARA\&A, 18, 165
\bibitem[1981]{mil81} Miley G. K., Heckman T. M., Butcher H. R., van Breugel W. J. M., 1981, \apj, 247, L5
\bibitem[1997]{neu97} Neumann M., Meisenheimer K., R\"oser H.-J., Fink H. H., 1997 \aap, 318, 383
\bibitem[1997]{nil97} Nilsson K., Heidt J., Pursimo T., Sillanpaeae A., Takalo L. O., Jaeger K., 1997, \apj, 484, L107
\bibitem[2000]{ost00} Ostrowski M., 2000, \mnras, 312, 579
\bibitem[2002]{ost02} Ostrowski M., Bednarz J., 2002, \aap, 394, 1141
\bibitem[1997]{ost97} Ostrowski M., Sikora M., Madejski G., Begelman M., eds., 1997, Relativistic Jets in AGNs, Krak\'ow
\bibitem[1989]{owe89} Owen F. N., Hardee P. E., Cornwell T. J., 1989, \apj, 340, 698
\bibitem[2003]{par03} Parma P., de Ruiter H. R., Capetti A., Fanti R., Morganti R., Bondi M., Laing R. A., Canvin J. R., 2003, \aap, 397, 127
\bibitem[1976]{pac76} Pacholczyk A. G., Scott J. S., 1976, \apj, 203, 313
\bibitem[1999]{per99} Perlman E. S., Biretta J. A., Zhou F., Sparks W. B., Macchetto F. D., 1999, \aj, 117, 2185
\bibitem[2001]{per01} Perlman E. S., Biretta J. A., Sparks W. B., Macchetto F. D., Leahy J. P., 2001, \apj, 551, 206
\bibitem[2001]{pes01} Pesce J. E., Sambruna R. M., Tavecchio F., Maraschi L., Cheung C. C., Urry C. M., Scarpa R., 2001, \apj, 556, L79
\bibitem[2003]{pfr03} Pfrommer C., Ensslin T. A., 2003, \aap, 407, 73
\bibitem[1991]{raw91} Rawlings S., Saunders R., 1991, Nature, 349, 138
\bibitem[1978]{ree78} Rees M. J., 1978, \mnras, 184, 61
\bibitem[2004]{rei04} Reimer A., Protheroe R. J., Donea A.-C., 2004, \aap, accepted (astro-ph/0402258)
\bibitem[1997]{rey97} Reynolds C. S., Begelman M. C., 1997, \apj, 487, 135
\bibitem[1997]{rid97} Ridgway S. E., Stockton A., 1997, \aj, 114, 511
\bibitem[1991]{ros91} R\"oser H.-J., Meisenheimer K., 1991, \aap, 252, 458
\bibitem[2000]{ros00} R\"oser H.-J., Meisenheimer K., Neumann M., Conway R. G., Perley R. A., 2000, \aap, 360, 99
\bibitem[1999]{row99} Rowell G. P., Dazeley S. A., Edwards P. G., Gunji S., Hara T., Holder J., Kawachi A., Kifune T., Matsubara Y., Mizumoto Y., Mori M., Muraishi H., Muraki Y., Naito T., Nishijima K., Ogio S., Patterson J. R., Roberts M. D., Sako T., Sakurazawa K., Susukita R., Tamura T., Tanimori T., Thornton G. J., Yanagita S., Yoshida T., Yoshikoshi T., 1999, APh, 11, 217
\bibitem[2003]{sah03} Sahayanathan S., Misra R., Kembhavi A. K., Kaul C. L., 2003, \apj, 588, 77
\bibitem[2001]{sam01} Sambruna R. M., Urry C. M., Tavecchio F., Maraschi L., Scarpa R., Chartas G., Muxlow T., 2001, \apj, 549, L161
\bibitem[2002]{sam02} Sambruna R. M., Maraschi L., Tavecchio F., Urry C. M., Cheung C. C., Chartas G., Scarpa R., Gambill J. K., 2002, \apj, 571, 206
\bibitem[2004a]{sam04a} Sambruna R. M., Gliozzi M., Donato D., Tavecchio F., Cheung C. C., Mushotzky R. F., 2004a, \aap, 414, 885
\bibitem[2004b]{sam04b} Sambruna R. M., Gambill J. K., Maraschi L., Tavecchio F., Cerutti R., Cheung C. C., Urry C. M., Chartas G., 2004b, \apj, accepted (astro-ph/0401475)
\bibitem[1983]{san83} Sanders R. H., 1983, \apj, 266, 73
\bibitem[2002]{sar02} Saripalli L., Subrahmanyan R., Udaya Shankar N., 2002, \apj, 565, 256
\bibitem[2003]{sar03} Saripalli L., Subrahmanyan R., Udaya Shankar N., 2003, \apj, 590, 181
\bibitem[1999]{sca99} Scarpa R., Urry C. M., Falomo R., Treves A., 1999, \apj, 526, 643
\bibitem[2002]{sca02} Scarpa R., Urry C. M., 2002, NewAR, 46, 405
\bibitem[1963]{sch63} Schmidt M., 1963, Nature, 197, 1040
\bibitem[2000]{scho00} Schoenmakers A. P., de Bruyn A. G., R\"ottgering H. J. A., van der Laan H., Kaiser C. R., 2000, \mnras, 315, 371
\bibitem[1979]{sch79} Schreier E. J., Feigelson E., Delvaille J., Giacconi R., Grindlay J. E., Schwartz D. A., Fabian A., 1979, \apj, 234, 39
\bibitem[1982]{sch82} Schreier E. J., Gorenstein P., Feigelson E. D., 1982, \apj, 261, 42
\bibitem[2002a]{sch02a} Schwartz D. A., 2002a, \apj, 569, L23
\bibitem[2002b]{sch02b} Schwartz D. A., 2002b, \apj, 571, L71
\bibitem[2000]{sch00} Schwartz D. A., Marshall H. L., Lovell J. E. J., Piner B. G., Tingay S. J., Birkinshaw M., Chartas G., Elvis M., Feigelson E. D., Ghosh K. K., Harris D. E., Hirabayashi H., Hooper E. J., Jauncey D. L., Lanzetta K. M., Mathur S., Preston R. A., Tucker W. H., Virani S., Wilkes B., Worrall D. M., 2000, \apj, 540, 69
\bibitem[2003]{sch03} Schwartz D. A., Marshall H. L., Miller B. P., Worrall D. M., Birkinshaw M., Lovell J. E. J., Jauncey D. L., Perlman E. S., Murphy D. W., Preston R. A., 2003, In: S. Collin, F. Combes and I. Shlosman, eds., Active Galactic Nuclei: from Central Engine to Host Galaxy, p. 359
\bibitem[2002]{sie02} Siemiginowska A., Bechtold J., Aldcroft T. L., Elvis M., Harris D. E., Dobrzycki A., 2002, \apj, 570, 543
\bibitem[2003a]{sie03a} Siemiginowska A., Stanghellini C., Brunetti G., Fiore F., Aldcroft T. L., Bechtold J., Elvis M., Murray S. S., Antonelli L. A., Colafrancesco S., 2003a, \apj, 595, 643
\bibitem[2003b]{sie03b} Siemiginowska A., Smith R. K., Aldcroft T. L., Schwartz D. A., Paerels F., Petric A. O., 2003b, \apj, 598, L15
\bibitem[1997]{sik97} Sikora M., Madejski G., Moderski R., Poutanen J., 1997, \apj, 484, 108
\bibitem[2000]{sik00} Sikora M., Madejski G., 2000, \apj, 534, 109
\bibitem[2002]{smi02} Smith D. A., Wilson A. S., Arnaud K. A., Terashima Y., Young A. J., 2002, \apj, 565, 195
\bibitem[1995]{spa95} Sparks W. B., Golombek D., Baum S. A., Biretta J., de Koff S., Macchetto F., McCarthy P., Miley G. K., 1995, \apj, 450, L55
\bibitem[2000]{spa00} Sparks W. B., Baum S. A., Biretta J., Macchetto F. D., Martel A. R., 2000, \apj, 542, 667
\bibitem[2002a]{sta02a} Stawarz \L ., Ostrowski M., 2002a, PASA, 19, 22
\bibitem[2002b]{sta02b} Stawarz \L ., Ostrowski M., 2002b, \apj, 578, 763
\bibitem[2003]{sta03} Stawarz \L ., Sikora M., Ostrowski M., 2003, \apj, 597, 597, 186
\bibitem[2004]{sta04} Stawarz \L ., Sikora M., Ostrowski M., Begelman, M. C., 2004, \apj, accepted (astro-ph/0401356)
\bibitem[1998]{swa98} Swain M. R., Bridle A. H., Baum S. A., 1998, \apj, 507, 29
\bibitem[2000]{tav00} Tavecchio F., Maraschi L., Sambruna R. M., Urry C. M., 2000, \apj, 544, L23
\bibitem[2003]{tav03} Tavecchio F., Ghisellini G., Celotti A., 2003, \aap, 403, 83
\bibitem[1995]{urr95} Urry C. M., Padovani P., 1995, PASP, 107, 803
\bibitem[1997]{war97} Wardle J. F. C., Aaron S. E., 1997, \mnras, 286, 425
\bibitem[2001]{wil01} Wilson A. S., Young A. J., Shopbell P. L., 2001, \apj, 547, 740
\bibitem[2002]{wil02} Wilson A. S., Yang Y., 2002, \apj, 568, 133
\bibitem[2001]{wor01} Worrall D. M., Birkinshaw M., Hardcastle M. J., 2001, \mnras, 326, L7
\bibitem[2003]{wor03} Worrall D. M., Birkinshaw M., Hardcastle M. J., 2003, \mnras, 343, 73
\end{thebibliography}
\end{document}